%% file: main.tex
\RequirePackage{lineno}
\documentclass[aps,superscriptaddress,showpacs,preprintnumbers,amsmath,amssymb]{revtex4}
\usepackage{graphicx,color,dcolumn,booktabs,bm}
\usepackage{longtable,lscape}
\usepackage{amsmath}
\usepackage{indentfirst}
\usepackage{epsfig}
\usepackage{feynmf}   
\usepackage{epstopdf}   
\usepackage{slashed}  
\usepackage{cases}
\usepackage{color}
\usepackage{multirow}
\usepackage{graphicx,color,dcolumn,booktabs,bm}
\usepackage{verbatim}
\usepackage[colorlinks,linkcolor=red,anchorcolor=green,citecolor=blue]{hyperref}
\newsavebox{\tablebox}
\usepackage{mathrsfs}
\usepackage{threeparttable}
\usepackage{lineno}
\usepackage[figuresright]{rotating}
 \usepackage[OT2,T1]{fontenc}
\usepackage{overpic}
\usepackage{graphicx}
\graphicspath{{./}}
\usepackage{ulem}

\usepackage{enumerate}
\usepackage{lineno}
\usepackage{graphicx} 
\usepackage{dcolumn}  

\usepackage[toc,titletoc]{appendix}
\def\yns{\Upsilon(nS)}

\def\Dst {D^{\ast+}}
\def\Dz {D^{0}}

\def\pip {\pi^{+}}
\def\pim {\pi^{-}}
\def\piz {\pi^{0}}

\def\Ks {K_S^0}
\def\Kp {K^{+}}
\def\Km {K^{-}}

\begin{document}


\title{Dalitz analysis of $\Dz\to\Km\pip\eta$ decays at Belle}

\input{author}

\begin{abstract}
We present the results of the first Dalitz plot analysis of the decay $\Dz\to\Km\pip\eta$. 
The analysis is performed on a data set corresponding to an integrated luminosity of 953 $\rm{fb}^{-1}$ 
collected by the Belle detector at the asymmetric-energy $e^{+}e^{-}$ KEKB collider. 
The Dalitz plot is well described by a combination of the six resonant decay channels $\bar{K}^{*}(892)^0\eta$, $\Km a_0(980)^+$, $\Km a_2(1320)^+$, $\bar{K}^{*}(1410)^0\eta$, $K^{*}(1680)^-\pip$ and  $K_2^{*}(1980)^-\pip$, together with  $K\pi$ and $K\eta$ S-wave components. 
The decays $K^{*}(1680)^{-}\to\Km\eta$ and $K_{2}^{*}(1980)^{-}\to\Km\eta$ are observed for the first time. 
We measure ratio of the branching fractions, $\frac{\mathcal{B}(D^0\to\Km\pip\eta)}{\mathcal{B}(D^0\to\Km\pip)}=0.500\pm0.002{\rm(stat)}\pm0.020{\rm(syst)}\pm0.003{\rm (\mathcal{B}_{PDG})}$. 
Using the Dalitz fit result, the ratio $\frac{\mathcal{B}(K^{*}(1680)\to K\eta)}{\mathcal{B}(K^{*}(1680)\to K\pi)}$ is measured to be $0.11\pm0.02{\rm(stat)}^{+0.06}_{-0.04}{\rm(syst)}\pm0.04{\rm(\mathcal{B}_{\text{PDG}})}$; this is 
 much lower than the theoretical expectations ($\approx1$) made under the assumption that $K^{*}(1680)$ is a pure $1^{3}D_1$ state. 
The product branching fraction $\mathcal{B}(D^0\to [K_2^{*}(1980)^-\to\Km\eta]\pip)=(2.2^{+1.7}_{-1.9})\times10^{-4}$ 
is determined. In addition, the $\pi\eta^{\prime}$ contribution to the $a_0(980)^{\pm}$ resonance shape is 
confirmed with 10.1$\sigma$ statistical significance using the three-channel Flatt\'{e} model. 
We also measure $\mathcal{B}(D^0\to\bar{K}^{*}(892)^0\eta)=(1.41^{+0.13}_{-0.12})\%$.
This is consistent with, and more precise than, the current world average $(1.02\pm0.30)\%$, deviates with a significance of more than $3\sigma$ 
from the theoretical predictions of (0.51-0.92)\%.

\end{abstract}

\pacs{13.25.Ft, 14.40.Lb}

\maketitle

\tighten

\section{\boldmath Introduction}
The understanding of hadronic charmed-meson decays is theoretically challenging due to the significant non-perturbative contributions, and input from experimental measurements 
thus plays an important role~\cite{Cheng:2010ry,Li:2012cfa,Li:2013xsa}. 
We present a Dalitz plot (DP) analysis~\cite{Dalitz} to study the dynamics of three body decay $D^0\to\Km\pip\eta$. This decay
is Cabibbo-favored (CF) and proceeds via the $c\to su\bar{d}$ transition. 
Because of isospin symmetry, intermediate states of this decay (e.g. excited kaon states decaying into $K\pi$ or $K\eta$), and $a$-family mesons decaying into $\pi\eta$, 
are similar to those in $\Dz\to\Ks\piz\eta$. 
The DP analysis of the latter channel has previously been performed, and the intermediate channels $K_S^0a_0(980)^0$ and $\bar{K}^{*}(892)^0\eta$~\cite{Rubin:2004cq} were found to be dominant,  
but additional components of a non-resonant amplitude, $K_{0}^{*}(1430)\eta$, $\Ks a_2(1320)$, $\kappa\eta$, and combinations of these processes, were found to contribute significantly.
However, the statistical power of that sample was too limited for precise measurements to be made. 
The $D^0\to \bar K^{*0}\eta$ decay is sensitive to the $W$-exchange diagram, which is important for the theoretical understanding of charm decays. 
The theoretical predictions of the branching fraction of this mode vary in the range (0.51-0.92)\% depending on the method~\cite{Cheng:2010ry,Li:2012cfa,Li:2013xsa}. 
This is consistent with, but smaller than, the current experimental result of $(1.02\pm0.30)\%$~\cite{PDG2018} obtained in the $D^0\to K_S^0\pi^0\eta$ final state~\cite{Rubin:2004cq}. 
A more precise measurement of this branching fraction from $D^0\to K^-\pi^+\eta$ decays would test the theoretical predictions. 
The $K^{*}_0(1430)^{\pm}\to K^{\pm}\eta$ decay was observed by the BABAR experiment~\cite{BABAR_Lees2014} and is awaiting confirmation. 
Experimentally, the ratio of $K^{*}_0(1430)$ decaying into $K\eta$ and $K\pi$ is $0.09^{+0.03}_{-0.04}$~\cite{PDG2018}, which is consistent with the theoretical prediction of 0.05~\cite{Barnes:2002mu} 
which was made with the assumption that it is a pure $1^{3}P_0$ state. 

Decays of some other excited kaons to $K\eta$, including $K^{*}(1410)$, $K^{*}(1680)$ and $K_2^{*}(1980)$, were predicted by Refs.~\cite{Barnes:2002mu,Pang:2017dlw} but have not yet been observed. 
These states may have some interesting properties; $K^*(1410)$ may not be a simple $2^3S_1$ state, and $K^*(1410)$ and $K^*(1680)$ are predicted to be a mixture of the $2^3S_1$ and $1^3D_1$ states. 
Assuming $K^*(1410)$ and $K^*(1680)$ are pure $2^3S_1$ and $1^3D_1$ states, respectively, the relative branching ratio of $K^*(1680)$ to $K\eta$ and $K\pi$ should be close to one (1.18 in Ref.~\cite{Barnes:2002mu} and 0.93 in Ref.~\cite{Pang:2017dlw}) Experimentally, no branching ratio measurement for the former channel has previously been made. 

The nature of the $a_0(980)$ is still not clear. Since it is a dominant intermediate resonance in $D^0\to\Km\pip\eta$, we can collect a large sample of $a_0(980)^+$ decays 
to study its character further, e.g. to confirm the $\pi\eta^{\prime}$ contribution to the $a_0(980)$ lineshape in a Flatt\'{e} model as measured by BESIII~\cite{bes3_a0}. 
Such a study can also help determine the $\pi\eta$ and $K\bar{K}$ contributions to $a_0(980)$ precisely and understand its quark component.

Wrong-sign (WS) decays play an important role in studies of $\Dz$-$\bar{D}^0$ mixing and $CP$ violation such as the first observation of $\Dz$-$\bar{D}^0$ mixing~\cite{ddbarmixngLHCb}. 
One possible mode for this, $\Dz\to\Kp\pim\eta$, will be reconstructed at Belle II, which aims at a data set fifty times~\cite{bib:BelleII} larger than that currently available from Belle. A time-dependent Dalitz analysis of this mode can be used to measure charm-mixing parameters, and for such a measurement an amplitude analysis of the right-sign decay, $\Dz\to\Km\pip\eta$, 
is needed to obtain the CF decay model.


This paper is organized as follows. Section~\ref{belledetector} briefly describes the Belle detector and data samples, and Sec.~\ref{selection} discusses event selection and parameterizations of signal and background and presents the measurement of the overall branching fraction. 
In Sec.~\ref{isobar}, we report the results of the DP analysis. 
The evaluation of the systematic uncertainties are discussed in Sec.~\ref{systematic}. 
Further study and discussion of the Dalitz fit results are presented in Sec.~\ref{sec:discussion}. Finally, the conclusions are presented in Sec.~\ref{conclusions}.

\section{Belle detector and Data sets}{\label{belledetector}} 
We perform a first Dalitz analysis of the decays $D^0\to\Km\pip\eta$~\cite{bib:conjugated} using $953~\rm fb^{-1}$ of data collected at or near the $\yns$ resonances (n=1, 2, 3, 4, 5), 
where 74\% of the sample is taken at the $\Upsilon(4S)$ peak, with the Belle detector~\cite{BelleDetector} operating at the KEKB asymmetric-energy $e^+e^-$ collider~\cite{KEKB}.
The Belle detector is a large-solid-angle magnetic spectrometer that consists of a silicon vertex detector (SVD), a 50-layer central drift chamber (CDC), an array of aerogel threshold Cherenkov counters (ACC), a barrel-like arrangement of time-of-flight scintillation counters (TOF), and an electromagnetic calorimeter comprised of CsI(Tl) crystals located inside a superconducting solenoid coil that provides a 1.5~T magnetic field. An iron flux-return located outside of the coil is instrumented to detect $K_L^0$ mesons and to identify muons. A detailed description of the Belle detector can be found elsewhere~\cite{BelleDetector}.

\section{\boldmath Event Selection and Yields}{\label{selection}} 
The signal decay chain consists of $\Dst\to\Dz\pi^+_s$ with $\Dz\to\Km\pip\eta$ and $\eta\to\gamma\gamma$; $D^{*}$ mesons are produced in $e^+e^-\to c\bar{c}$ processes, and the charge of the slow pion $\pi_s$ tags the flavor of the $D^0$ meson~\cite{whyTag}. To ensure charged tracks are well reconstructed, each is required to have at least two associated hits of the SVD in the 
beam and azimuthal directions, separately. 
The slow-pion candidates are required to have 
the signed distances from the pivotal point to the helix to be within $\pm1.0$ cm in the transverse plane and within $\pm3.0$ cm along the direction opposite to the positron beam.
A charged track is identified as a kaon by requiring a ratio of particle identification likelihoods~\cite{bib:PID} ${\mathcal{L}_K}/({\mathcal{L}_K+\mathcal{L}_{\pi}})>0.7$, 
which are constructed using CDC, TOF, and ACC information; otherwise the track is assumed to be a pion. This requirement has efficiencies of 85\% and 98\% and misidentification rates of 2\% and 10\% for kaons and pions, respectively.
The photon candidates are reconstructed from ECL clusters unmatched to any charged track. 
The ratios of their energy deposits in a 3$\times$3 array of CsI(Tl) crystals to that in a 5$\times$5 array centered on the crystal with maximum deposited energy are required to be more than 0.8. 
The energies of photon candidates used to form $\eta$, $E_{\gamma}$, must exceed 60 or 120 MeV in the barrel or endcap region. 
The $\eta$ candidates must have $\gamma\gamma$ mass within $^{-0.06}_{+0.05}$ GeV/$c^2$ of the nominal mass~\cite{PDG2018} which takes into account the asymmetric resolution, and to 
have momentum in the laboratory frame, $p_{\eta}$, larger than 1 GeV/$c$. Furthermore, we require $|\cos \theta_{\eta}|=|\frac{E_{\gamma1}-E_{\gamma2}}{E_{\gamma1}+E_{\gamma2}}|\cdot\frac{E_{\eta}}{p_{\eta}}$ to be less than 0.8, which is optimized to suppress combinatorial background.
A large set of simulated signal Monte Carlo (MC) samples, more than 25 million events, is produced to study the efficiency. These are generated uniformly in phase space with the {\scshape{Evtgen}}~\cite{evtgen} and {\scshape{Jetset}}~\cite{jetset} software packages, and the detector response is modeled by the {\scshape{Geant3}}~\cite{geant3}. 
The final-state radiation effect is taken into account using the {\scshape{Photos}}~\cite{photons} package. 

Kaon and pion tracks with opposite charge are required to form a common vertex (the $D^0$ decay position) with fit quality $\chi^2_v$. 
The $\eta$ candidates are mass-constrained assuming that they are produced at this decay vertex, and the resultant $\eta$ momentum is added to the $K\pi$ system to obtain the $D^0$ momentum. 
The invariant mass of $K\pi\eta$, $M$, is required to satisfy the condition $1.80~{\rm GeV}/c^2< M < 1.92~{\rm GeV}/c^2$. Then, the $D^0$ production vertex is constrained to the $e^+e^-$ interaction point, with fit quality $\chi^2_b$. The $\pi_s$ track is refit to this $D^0$ production vertex, with a fit quality denoted $\chi^2_s$, to improve 
the resolution of the released energy in $\Dst$ decay, $Q\equiv M_{K\pi\eta\pi_s}-M_{K\pi\eta}-m_{\pi_s}$. 
The value of $Q$ is required to be less than 15 MeV/$c^2$ to suppress further combinatorial background. 
The $D^{*}$ momentum in the center-of-mass frame, $p^{\ast}$($D^{*}$), is required to be greater than 2.4, 2.5, or 3.1 GeV/$c$ for data below, on, or above $\Upsilon$(4S) energy, to reduce high-multiplicity events and combinatorial background. A consequence of this requirement is that the $D^0$ candidates from $B$ decays are removed. 
After applying all of these selection criteria, there are on average 1.3 signal decay candidates per event. A best-candidate selection (BCS) method is applied to multi-candidate events, retaining as the best candidate the one with the smallest sum of vertex-fit qualities, $\chi^2_v+\chi^2_b+\chi^2_s$. 
A mass-constrained fit is then applied to the $\Dz$ meson to improve resolution on the Dalitz variables, 
$M_{K\pi}^2$ and $M_{\pi\eta}^2$.

To extract yields of signal and background, a fit of the two-dimensional distribution of $M$ and $Q$ is performed. 
For the signal, the probability density function (PDF) in $M$ is described by the sum of a double Gaussian and a double bifurcated Gaussian, with a common mean value ($\mu$); the PDF in $Q$ is described by the sum of a bifurcated Student function, a bifurcated Gaussian function, and a bifurcated Cruijff function~\cite{Cruijff}, 
where the mean values and widths are correlated to the $M$ value by a second-order polynomial function of $|M-\mu|$. 
For a real signal $D^0$ combined with a random $\pi_s$ (named the random $\pi_s$ background), the $M$ distribution uses the same PDF as for the signal and the $Q$ distribution uses a threshold function, $f(Q)=Q^{\alpha}e^{-\beta Q}$. 
This random background will be treated as signal, as it nearly consists of the same $D^0$ decay as the signal when the tiny fraction of DCS decay relative to CF decay is neglected.
The combinatorial background is considered to have two components. A PDF smoothed by bilinear interpolation~\cite{bilinear} is used for correlated combinatorial background, 
which has a correctly reconstructed $\pi_s$ from $D^*$ decay, but incorrectly reconstructed $D^0$, whereas for other combinatorial background a third-order polynomial function of $M$ and a threshold function of $Q$ is used as a parameterization. 
The ratio between these two combinatorial backgrounds is fixed to that found using the generic MC.
Figure~1 shows the $M$-$Q$ combined fit for the experimental data. 
We obtain a signal yield of $105\,197\pm990$ in the $M$ and $Q$ two-dimensional (2D) signal region of $1.85~{\rm GeV}/c^2 < M < 1.88~{\rm GeV}/c^2$ and $5.35~{\rm MeV}/c^2 < Q < 6.35~{\rm MeV}/c^2$ with a high purity $(94.6\pm0.9)\%$. These are the combinations that will be used for the fit to the Dalitz plot.

\begin{figure}[!hbtp]
    \begin{centering}
    \includegraphics[width=0.80\textwidth]{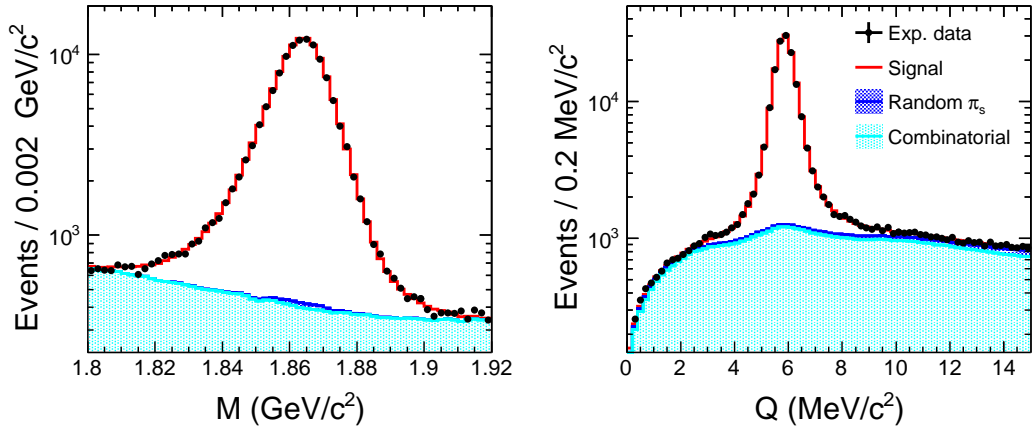}
    \caption{\label{fig:real_mqfit}The $D^0\to\Km\pip\eta$ reconstructed mass, $M$ (in $5.35~{\rm GeV}/c^2 < Q < 6.35~{\rm GeV}/c^2$) and release energy of $D^{*}$ decay, $Q$ (in $1.85~{\rm GeV}/c^2 < M < 1.88~{\rm GeV}/c^2$) for experimental data (points with error bars) and fitted contributions of signal, random $\pi_s$ and combinatorial backgrounds. }
    \end{centering}
\end{figure}
 
 To measure the branching fraction of the decay $D^0\to\Km\pip\eta$, we normalize the signal yield by the number of $D^0$ mesons produced in the decay $D^{*+}\to D^0\pi^+$. For normalization, we choose the $D^0\to\Km\pip$ channel, which has a well-known rate of $\mathcal{B}=(3.950\pm0.031)\%$~\cite{PDG2018}. 
 We use the same selection criteria as are used $D^0\to\Km\pip\eta$ but without the $\eta$. We extract the signal yield from the distribution of $D^0$ invariant mass in $1.78~{\rm GeV}/c^2<M<1.94~{\rm GeV}/c^2$ and $Q$ wide signal region $|Q-5.85|<1.0$ MeV/$c^2$ and find signal yields of $116\,302\pm510$ for $D^0\to\Km\pip\eta$, and $2\,597\,343\pm1\,669$ for $D^0\to\Km\pip$ (with a high purity 98.3\%) based on the $\Upsilon$(4S) on-resonance data set. 
 The efficiency $\epsilon(D^0\to\Km\pip\eta)=(5.34\pm0.01)\%$ and $\epsilon(D^0\to\Km\pip)=(23.49\pm0.02)\%$ are determined based on Dalitz signal MC produced with the nominal Dalitz fit result shown in Table~\ref{tab:DPfitexp} for $D^0\to\Km\pip\eta$ and signal MC for $D^0\to\Km\pip$. 
Taking into account the branching fraction $\mathcal{B}(\eta\to\gamma\gamma)=(39.41\pm0.20)\%$~\cite{PDG2018}, we find the ratio of branching fractions to be 
 \begin{eqnarray}
 \frac{\mathcal{B}(D^0\to\Km\pip\eta)}{\mathcal{B}(D^0\to\Km\pip)}=0.500\pm0.002{\rm(stat)}\pm0.020{\rm(syst)}\pm0.003{\rm (\mathcal{B}_{PDG})},
 \end{eqnarray}
 where the three uncertainties shown are statistical, systematic, and the uncertainty of branching fraction of $\eta\to\gamma\gamma$, respectively. 
 Using the known $D^0\to\Km\pip$ branching fraction, we measure the branching fraction
 \begin{eqnarray}
 \mathcal{B}(D^0\to\Km\pip\eta)=(1.973\pm0.009{\rm(stat)}\pm0.079{\rm(syst)}\pm0.018{\rm (\mathcal{B}_{PDG})})\%,
 \end{eqnarray}
where the last error is associated with uncertainty of the branching fractions of $D^0\to\Km\pip$ and $\eta\to\gamma\gamma$. Many systematic uncertainties are canceled in the ratio measurement, 
and the dominant uncertainty is that of the $\eta$ reconstruction efficiency (4\%). 

\section{\boldmath Dalitz analysis}{\label{isobar}} 
The isobar model~\cite{isobar model} is applied for the amplitude of $\Dz\to (R\to AB)C$ through a resonance $R$ with spin-$J$ ($A$, $B$ and $C$ are pseudoscalar particles). The decay amplitude is given by a coherent sum of individual contributions, consisting of a constant term $a_{NR}e^{i\phi_{NR}}$ for the non-resonant three-body decay, and different quasi-two-body resonant decays: 
\begin{eqnarray}
{\mathcal M}=a_{NR}e^{i\phi_{NR}} + \sum_{R}a_Re^{i\phi_R}{\mathcal{M}_R}(m_{AB}^2, m_{BC}^2). \label{eqn:isobar}  
\end{eqnarray}
Here $m^2_{AB}$ and $m^2_{BC}$ are Dalitz variables, and $a_Re^{i\phi_R}$ is a complex amplitude for the contribution of an individual intermediate resonance $R$. The amplitude and phase of $\bar{K}^*(892)^0$, having the largest fit fraction, are fixed to $a_{K^*(892)^0}=1$ and $\phi_{K^*(892)^0}=0$. The matrix element $\mathcal{M}_R$ for an intermediate resonant decay is given by
\begin{eqnarray}
\mathcal{M}(ABC|R)=F_D\times F_R\times T_R \times\Omega_J,
\end{eqnarray}
where $T_R \times\Omega_J$ is a resonance propagator. $T_R$ is a dynamical function for a resonance, described by a relativistic Breit-Wigner (RBW) with mass-dependent width,
\begin{eqnarray}
T_R = \frac{1}{M_R^2 - m_{AB}^2 - iM_R\Gamma_{AB}}, \quad \quad 
\Gamma_{AB} = \Gamma^R_0 \left ( \frac{p_{AB}}{p_R}\right )^{2J+1}\left ( \frac{M_{R}}{m_{AB}}\right ) F_R^2, \label{eqn:dynamic}\end{eqnarray}
where $p_{AB}$ ($p_{R}$) is the momentum of either daughter in the $AB$ (or $R$) rest frame, and $M_R$ and $\Gamma^R_0$ are the nominal mass and width, 
$\Omega_J$ describes the angular momentum that depends on the spin $J$ by using the Zemach tensor~\cite{Zemach,Kpipi0_CLEO}, 
and $F_D$ and $F_R$ are Blatt-Weisskopf centrifugal barrier factors~\cite{bib:BWfactor,bib:BWfactor2}, describing the quark structure of the $\Dz$ meson and intermediate resonance. 
The parameter of meson radius, $R$, is set to 5.0 (GeV/$c$)$^{-1}$ and 1.5 (GeV/$c$)$^{-1}$ for the $\Dz$ meson and the intermediate resonances, respectively~\cite{Kpipi0_CLEO}.
For the $a_0(980)$ contribution description, we use the Flatt\'e formalism with three coupled channels, $\pi\eta$, $\bar{K}^0K$ and $\pi\eta'$~\cite{bes3_a0}
\begin{equation}
 T_R(s)=\frac{1}{m_{a_0}^2-s-i(g^2_{\pi\eta}\rho_{\pi\eta}+g^2_{\bar{K}^0K}\rho_{\bar{K}^0K}+g^2_{\pi\eta'}\rho_{\pi\eta'})}, \label{eqn:a09803}
\end{equation}
where $\sqrt{s}$ is the invariant mass of $\pi\eta$; $g_i$ and $\rho_i$ are coupling constants and phase-space factors, respectively. For example  
 $\rho_{\pi\eta} = \sqrt{\left[1-(m_\pi+m_\eta)^2/M_{\pi\eta}^2\right]\left[1-(m_\pi-m_\eta)^2/M_{\pi\eta}^2\right]}$.

The generalized LASS model~\cite{PBF,LASS2} is used to parameterize the $K\pi$ and $K\eta$ S-wave contributions:
\begin{eqnarray}
\mathcal{A}_{gLASS}(s)=\frac{\sqrt{s}}{2q}\cdot [B\sin(\delta_B+\phi_B)e^{i(\delta_B+\phi_B)} + \sin(\delta_R)e^{i(\delta_R+\phi_R)}e^{2i(\delta_B+\phi_B)}], \label{eqn:lass}
\end{eqnarray}
where $s$ is the invariant mass squared of the $K\pi$ or $K\eta$ system, $q$ is the momentum of $K$ in the $K\pi$ or $K\eta$ rest frame, and $\delta_B$ and $\delta_R$ are phase angles of the non-resonant component and $K_0^{*}(1430)$ component, respectively. They are defined as
$\tan(\delta_R) = M_r \Gamma(m_{ab})/(M_r^2 - m^2_{ab})$ and $\cot(\delta_B) = 1/(aq) + rq/2$,
where $a$, $r$, $B$, $\phi_B$ and $\phi_R$ are real parameters and may be determined by amplitude analysis. 

The DP fit is performed by an unbinned maximum likelihood method with
\begin{eqnarray}
\ln\mathcal{L} & = & \sum\limits_{i=1}^{n}\ln[f_{s}^i(M_i,Q_i)\cdot P_{s}(m_{K\pi,i}^2,m_{\pi\eta,i}^2) + (1-f_{s}^i(M_i,Q_i))\cdot P_{b}(m_{K\pi,i}^2,m_{\pi\eta,i}^2)] ,
\end{eqnarray}
where $n$ is the number of $\Dz$ candidates in the $M$ and $Q$ 2D signal region and $f_{s}^i$ is the event-by-event fraction of signal obtained from the $M$-$Q$ fit; 
the combinatorial background function, $P_{b}$, is a smoothed PDF~\cite{bilinear}, determined from the DP in the $M$ sideband region ($1.755~{\rm GeV}/c^2 < M < 1.775~{\rm GeV}/c^2$ or $1.935~{\rm GeV}/c^2 < M < 1.955~{\rm GeV}/c^2$) and the $Q$ signal region ($5.35~{\rm MeV}/c^2 < Q < 6.35~{\rm MeV}/c^2$). 
The signal PDF, $P_s$, is calculated taking the reconstruction-efficiency dependence on the Dalitz-plot variables into account, and normalized in the Dalitz plot region.
\begin{eqnarray}
P_{s}=|\mathcal{M}(m_{K\pi}^2,m_{\pi\eta}^2)|^2\epsilon(m_{K\pi}^2,m_{\pi\eta}^2)/\iint dm_{K\pi}^2dm_{\pi\eta}^2|\mathcal{M}(m_{K\pi}^2,m_{\pi\eta}^2)|^2\epsilon(m_{K\pi}^2,m_{\pi\eta}^2).
\end{eqnarray}
 This efficiency distribution $\epsilon(m_{K\pi,i}^2,m_{\pi\eta,i}^2)$ is obtained from a high-statistic signal-MC sample and takes into account the known difference in particle identification efficiency for charged tracks between MC and data. These correction factors depend on the momentum and polar angle of individual charged track.  
The fit fractions (FF) of each intermediate component are calculated across the DP region as 
\begin{eqnarray}
FF = \dfrac{\iint_{DP} |a_R e^{i\phi_R} {\mathcal M}_{R}(m_{K\pi}^2,m_{\pi\eta}^2)|^2 dm_{K\pi}^2dm_{\pi\eta}^2}{\iint_{DP} |{\mathcal M}(m_{K\pi}^2,m_{\pi\eta}^2)|^2 dm_{K\pi}^2dm_{\pi\eta}^2}.   \label{eqn:fitfrac}
\end{eqnarray}
The FF uncertainties are evaluated using a Toy MC method in which the sampling takes into account the considerations among all the fitted parameters by propagating the full covariance matrix obtained by the DP fit.

\begin{figure}[!htpb]
  \begin{overpic}[width=0.4\textwidth]{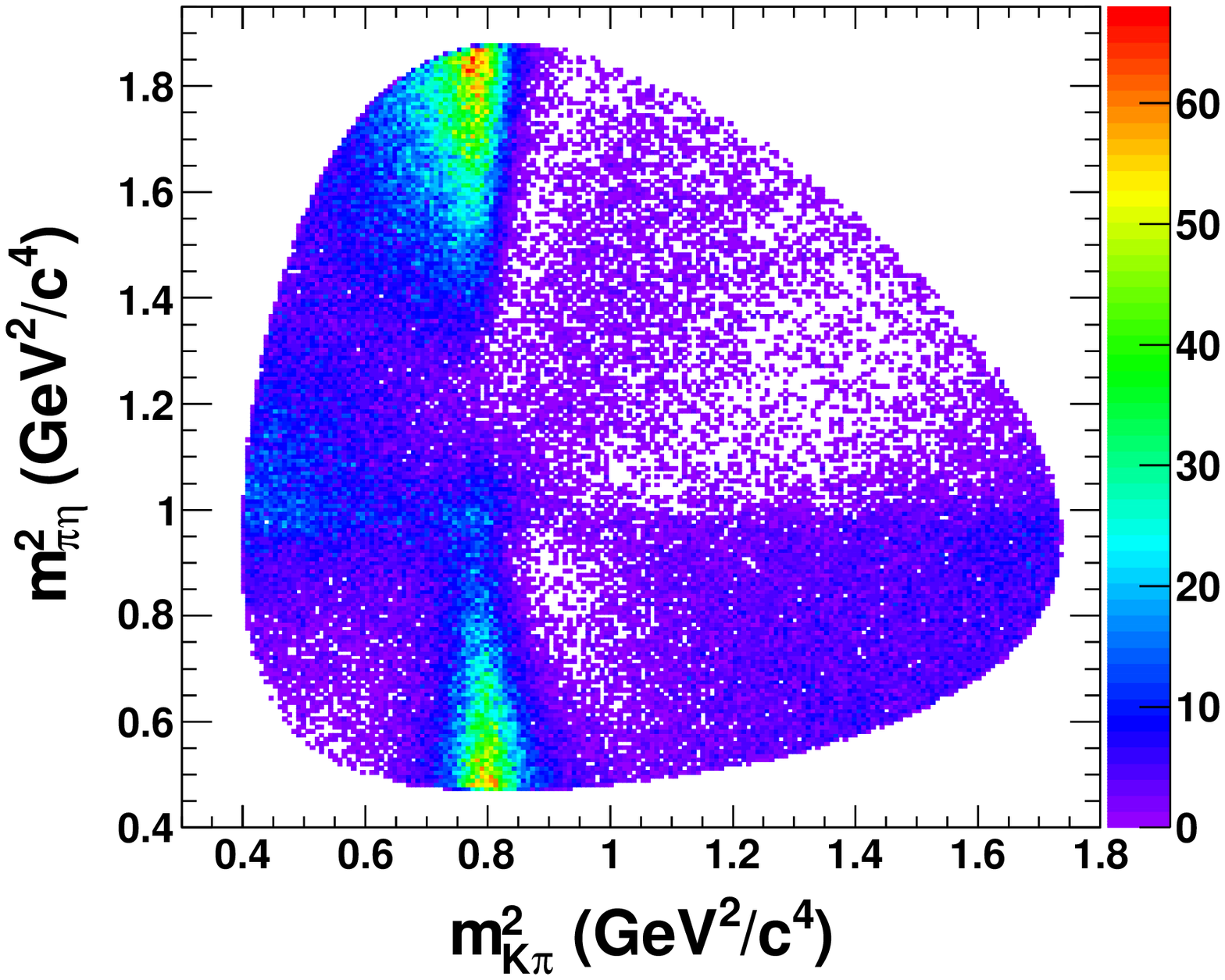}
  \put(21, 65){\large(a)}
  \end{overpic}%
  \begin{overpic}[width=0.4\textwidth]{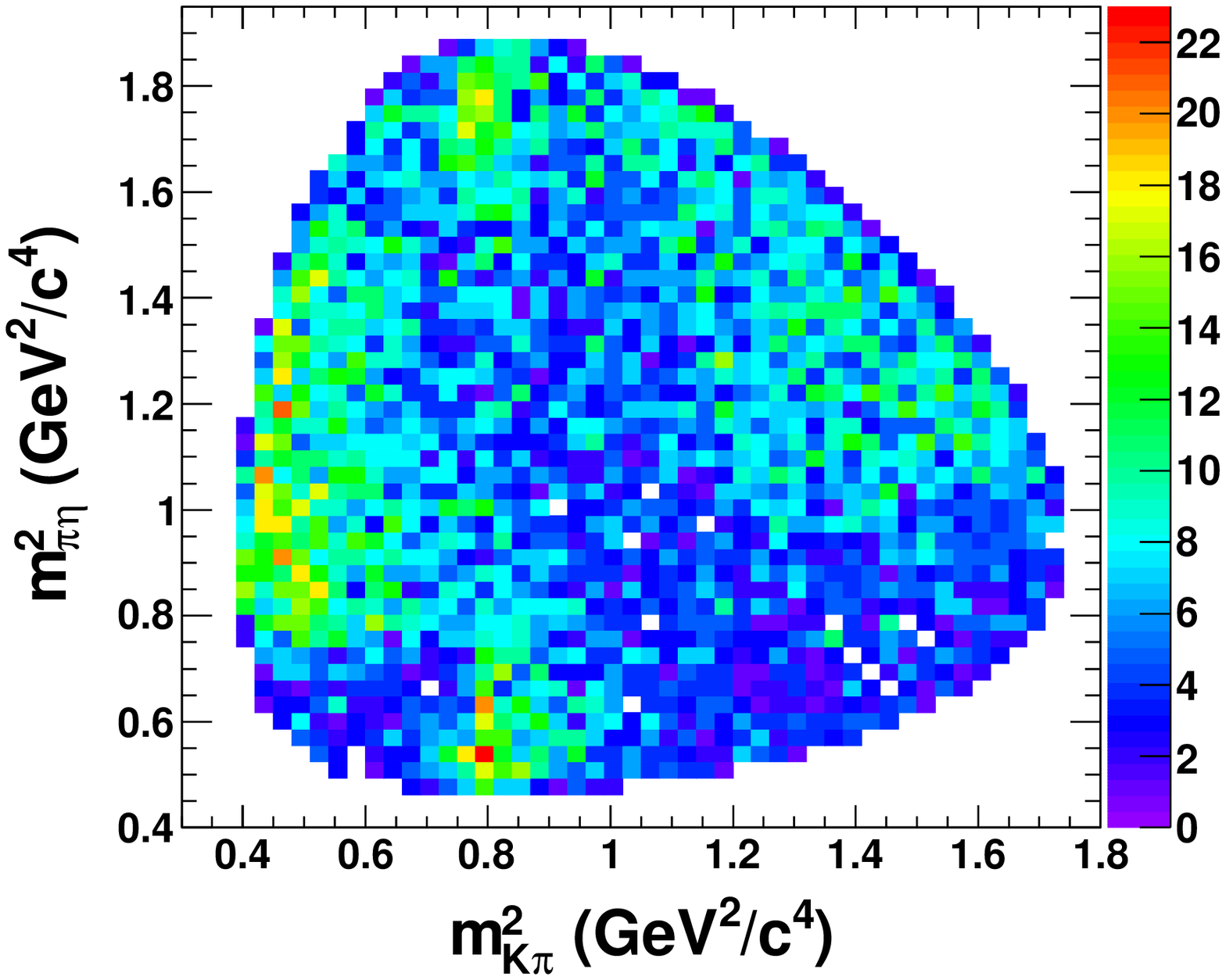}
  \put(21, 65){\large(b)}
  \end{overpic}\\
  \begin{overpic}[width=0.33\textwidth]{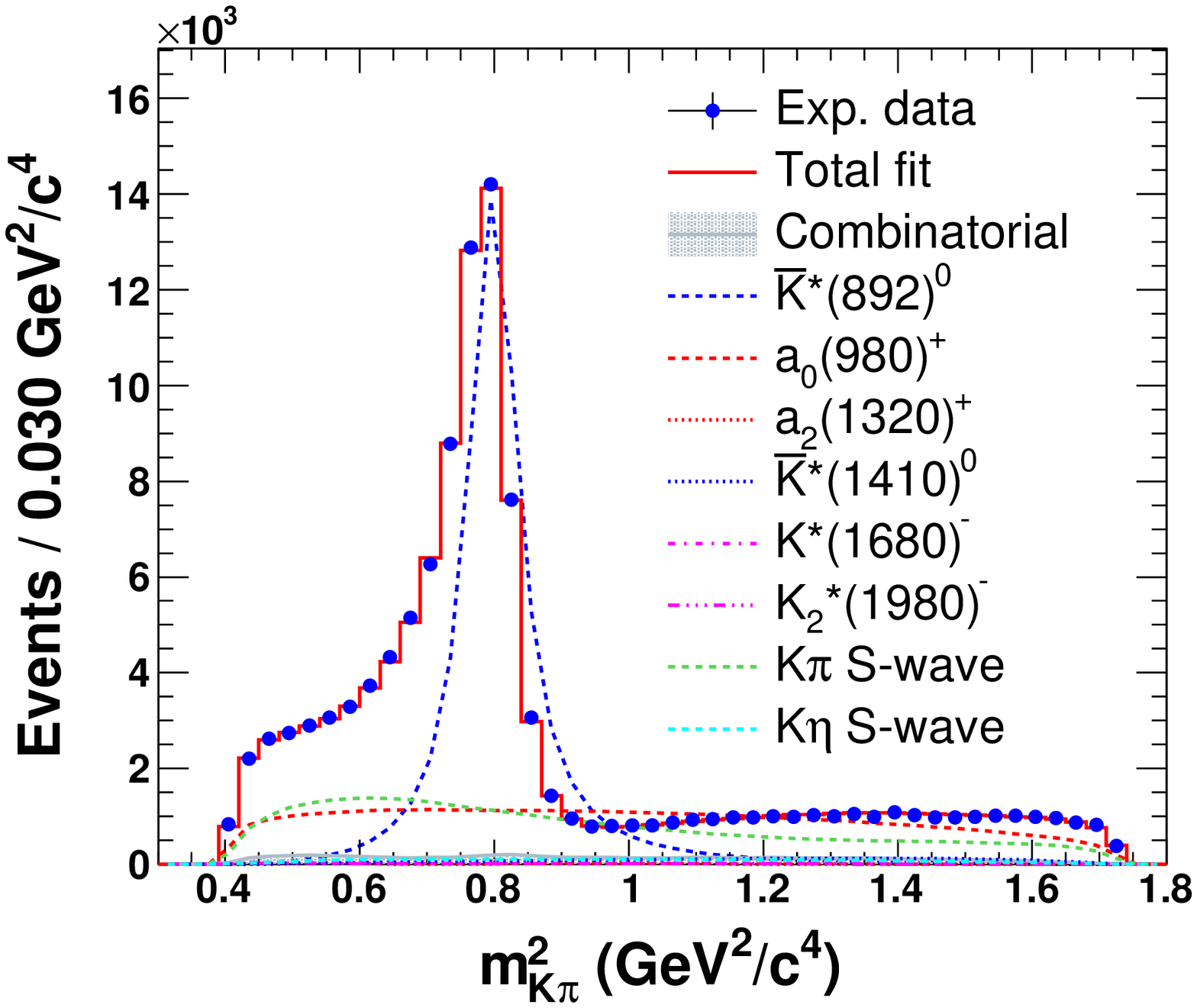}
  \put(22, 68){\large(c)}
  \end{overpic}%
  \begin{overpic}[width=0.33\textwidth]{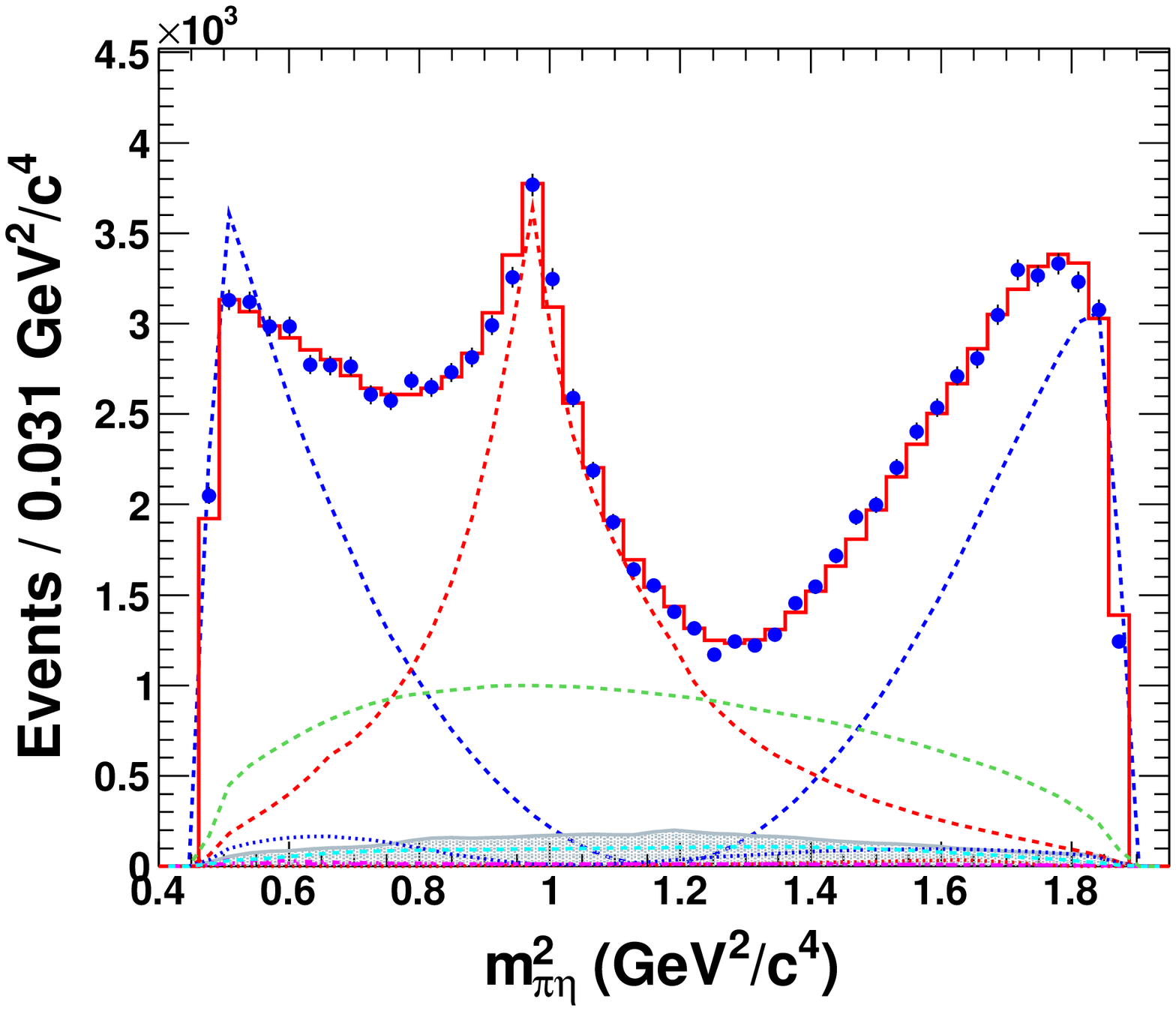}
  \put(22, 68){\large(d)}
  \end{overpic}%
  \begin{overpic}[width=0.33\textwidth]{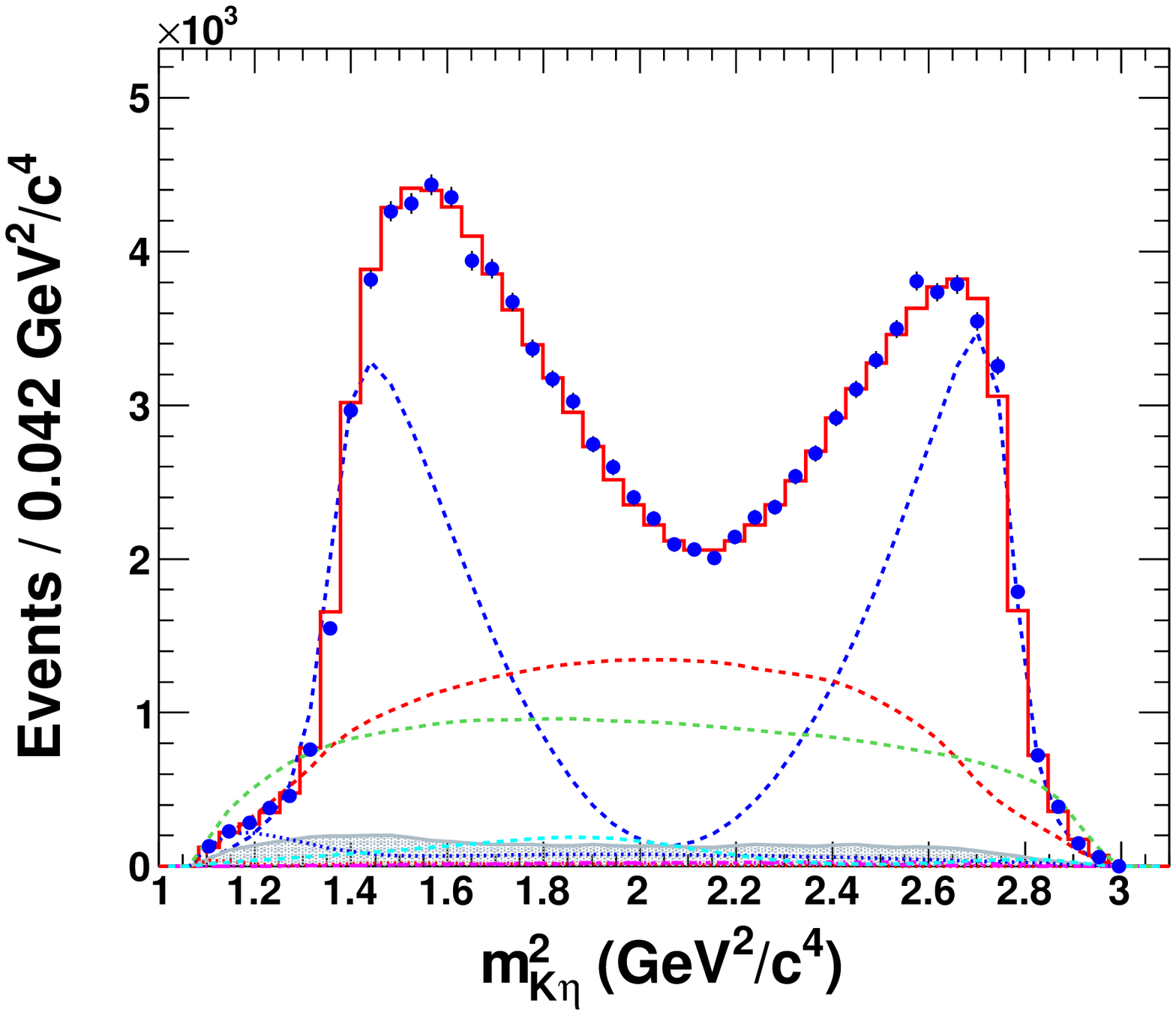}
  \put(22, 68){\large(e)}
  \end{overpic}%
  \vskip-15pt
  \caption{\label{fig:10res} The Dalitz plot of $\Dz\to\Km\pip\eta$ in (a) $M$-$Q$ signal region and (b) $M$ sideband region, and projections on (c) $m_{K\pi}^2$, (d) $m_{\pi\eta}^2$ and (e) $m_{K\eta}^2$. In projections the fitted contributions of individual components are shown, along with contribution of combinatorial background (grey-filled) from sideband region.}
\end{figure}

Fifteen possible intermediate resonances~\cite{bib:resonances} were initially considered in the Dalitz analysis. 
We found $\bar{K}^*(1410)^0$ and $\bar{K}^*(1680)^0$ have a phase-angle difference of approximately $180^{\circ}$ 
and similar behavior in the DP, therefore, it is hard to separate them. 
In order to ensure stability of the fit to the Dalitz plot, only $\bar{K}^*(1410)^0$ is kept, while a possible $\bar{K}^{*}(1680)^0$ contribution is considered as a source of systematic uncertainty. 
Therefore for the rest of this paper, $K^{*}(1410)^0$ represents the contribution of $K^{*}(1410)^0$, $K^{*}(1680)^0$ and their possible interference.
Then, the resonances not contributing to the amplitude significantly are eliminated one by one based on significance-level testing. 
Significances of individual contributions are determined as the likelihood difference, $\Delta(-2\ln\mathcal{L})$, that arises
when an individual contribution is removed from the model taking into account the degrees of freedom (d.o.f). 
Only components with significances in excess of $5\sigma$, i.e. $\Delta(-2\ln\mathcal{L})>28.74$ with $\Delta(\rm{d.o.f})=2$, are retained in the Dalitz model. Of the resonances which were eliminated, $K_2^{*}(1430)^-$ had the largest significance ($3.8\sigma$). 
A model with eight components is chosen as our nominal model, and this is presented in Fig.~\ref{fig:10res}. 
It includes six resonances [$a_0(980)^{+}$, $a_2(1320)^{+}$, $\bar{K}^{*}(892)^{0}$, $\bar{K}^{*}(1410)^{0}$, $K^*(1680)^{-}$, $K^{*}_2(1980)^{-}$] and two S-wave components ($K\pi$ and $K\eta$). 
The fit quality of this nominal model is $\chi^2/\mathrm{d.o.f}=1638/(1415-24)=1.18$ across the Dalitz plane, and the three Dalitz plot projections are shown in Fig.~\ref{fig:10res} (b-d). 
The statistical significance of each component is larger than $10\sigma$. 
In particular, the statistical significance of the $K\eta$ S-wave component with $K_0^{*}(1430)^{-}$ is greater than $30\sigma$, and $K^{*}(1680)^{-}\to\Km\eta$ and $K^{*}_2(1980)^{-}\to\Km\eta$ are observed for the first time and have statistical significances of $16\sigma$ and $17\sigma$, respectively. 
The fitted magnitudes and phases of intermediate components are listed in Tab.~\ref{tab:DPfitexp}, together with corresponding fit fractions, where statistical uncertainties are obtained from 500 sets of toy MC samples, 
and systematic uncertainties take into account model uncertainties and other systematic uncertainties as discussed in Sec.~\ref{systematic}. 
The fact that the sum of fit fractions is greater than 100\% indicates significant destructive interference.
Table~\ref{tab:DPfitexppar} shows the fitted parameters of LASS model in Eq.(\ref{eqn:lass}) and their correlation coefficient matrix for the $K\pi$ and $K\eta$ S-wave components. 
he left coefficients in the full correlation matrix from Dalitz fit are shown in Tab.~\ref{tab:corMatrix} including the correlation coefficients among the magnitudes and phases of resonances and the LASS model parameters.
Various Dalitz models, including the nominal model used in the fit to final experimental data, are produced using MC to perform tests for any possible bias, and to check that the input and output Dalitz parameters are consistent. 
We also checked for the existence of possible multiple solutions in the fit, with likelihood scanning of each of the free parameters. In addition, 100 sets of Dalitz fits were performed by sampling the initial values of free parameters uniformly in an interval around their final values. No multiple solutions were found. 

\begin{table}[!htbp]
\caption{\label{tab:DPfitexp}Magnitude and phase of intermediate components, and their fit fraction from Dalitz-plot fit of $D^0\to\Km\pip\eta$. The quoted uncertainties on the fit fractions are statistical, systematic, and the uncertainty due to the Dalitz model, respectively. } 
\begin{tabular}{c|ccc}
  \hline \hline
  Component            	&   Magnitude    	& Phase ($^\circ$)   	&   Fit fraction ($\%$)    \\ \hline
  $\bar{K}^{*}(892)^{0}$   &  $1$          	&  $0$   				& $47.61\pm1.32^{+0.24+3.64}_{-0.49-2.71}$ \\  
  ${a_{0}(980)^{+}}$       	&  $2.779\pm0.032$ &  $310.3\pm1.1$	& $39.28\pm1.50^{+1.58+4.38}_{-0.51-3.30}$ \\ 
  $K\pi$ S-wave	   	& $10.82\pm0.23$	&  $50.0\pm5.7$	& $31.92\pm1.21^{+1.47+2.75}_{-0.53-2.87}$ \\
  $K\eta$ S-wave   	 	&  $1.70\pm0.082$ 	&  $113.8\pm13.6$	& $3.37\pm0.50^{+0.77+3.20}_{-0.27-1.21}$  \\
  $a_{2}(1320)^{+}$		&  $1.27\pm0.079$ 	&  $283.4\pm4.7$	& $0.74\pm0.09^{+0.06+0.37}_{-0.04-0.17}$  \\  
  $\bar{K}^{*}(1410)^{0}$  	&  $4.84\pm0.36$ 	&  $352.7\pm2.8$	& $6.94\pm0.85^{+0.55+2.37}_{-1.61-3.22}$  \\
  $K^{*}(1680)^{-}$  		&  $2.56\pm0.18$ 	&  $232.2\pm6.6$	& $1.07\pm0.16^{+0.11+0.58}_{-0.10-0.36}$  \\ 
  ${K_{2}^{*}(1980)^{-}}$   	&  $9.29\pm0.69$ 	&  $207.7\pm4.0$	& $1.13\pm0.15^{+0.05+0.88}_{-0.05-0.98}$  \\  
  Sum			&       	   		&					& $132.1\pm3.4^{+1.6+8.3}_{-0.7-4.5}$  \\ 
   \hline\hline     
\end{tabular}
\end{table}

\begin{table}[!htbp]
\begin{center}    
\caption{\label{tab:DPfitexppar}Fitted parameters of the LASS model (with statistical uncertainties only) and their correlation coefficient matrix for $K\pi$ and $K\eta$ S-wave components.}            
\begin{tabular}{cr@{$\pm$}lrrrrrrrrrr} \hline \hline     
\multirow{2}{*}{Parameters} & \multicolumn{2}{c}{\multirow{2}{*}{fitted values}} & \multicolumn{10}{c}{correlation coefficient} \\ \cline{4-13} 
            &    \multicolumn{2}{c}{}  & $B^{K\pi}$ & $\phi_B^{K\pi}$ & $\phi_R^{K\pi}$ & $a^{K\pi}$ & $r^{K\pi}$ & $B^{K\eta}$ & $\phi_B^{K\eta}$ & $\phi_R^{K\eta}$ & $a^{K\eta}$ & $r^{K\eta}$ \\ \hline
$B^{K\pi}$ & $0.239$ & $0.010$ & $1$ & & & & & & & & &  \\
$\phi_B^{K\pi}$ ($^{\circ}$) & $-2.1$ & $0.8$ & $0.094$  & $1$   & & & & & & & & \\
$\phi_R^{K\pi}$ ($^{\circ}$) & $-0.7$ & $1.8$ & $0.134$  & $0.738$ & $1$  & & & & & & &  \\
$a^{K\pi}$ (GeV$^{-1}c$)  & $5.36$ & $0.29$ & $0.172$  & $0.784$ & $0.754$ & $1$   & & & & & & \\
$r^{K\pi}$ (GeV$^{-1}c$) & $-3.30$ & $0.10$ & $-0.385$ & $0.484$ & $0.409$ & $0.452$ & $1$  & & & & & \\
$B^{K\eta}$ & $0.693$ & $0.108$ &$-0.021$ & $0.309$ & $0.351$ & $0.278$ & $-0.185$ & $1$  & & & &  \\
$\phi_B^{K\eta}$ ($^{\circ}$) & $1.3$ & $3.4$ & $-0.318$ & $0.387$ & $0.340$  & $0.432$ & $0.529$  & $-0.338$ &$1$  & & &   \\
$\phi_R^{K\eta}$ ($^{\circ}$) & $25.5$ & $9.1$ & $-0.210$  & $-0.746$ & $-0.756$ & $-0.804$ & $-0.250$  & $-0.199$ & $-0.447$ & $1$  & &  \\
$a^{K\eta}$ (GeV$^{-1}c$) & $0.293$ & $0.048$ & $-0.373$ & $-0.711$ & $-0.696$ & $-0.790$  & $-0.214$ & $-0.173$ & $-0.509$ & $0.784$ & $1$  & \\
$r^{K\eta}$ (GeV$^{-1}c$)  & $-15.9$ & $2.6$ & $-0.381$ & $-0.694$ & $-0.675$ & $-0.776$ & $-0.218$ & $-0.092$ & $-0.528$ & $0.774$ & $0.995$ & $1$  \\ \hline \hline     
\end{tabular}     
\end{center}      
\end{table}

\begin{table}
\begin{centering}
\caption{\label{tab:corMatrix}The correlation coefficients among the resonant parameters: the magnitudes (mag.) and phases, and the LASS model parameters for $K\pi$ and $K\eta$ S-wave components from Dalitz fit.}
\begin{lrbox}{\tablebox}
\begin{tabular}{ccrrrrrrrrrrrrrr} \hline \hline
\multicolumn{2}{c}{\multirow{2}{*}{Parameters}} & \multicolumn{2}{c}{$a_0(980)^+$} & \multicolumn{2}{c}{$a_2(1320)^+$} & \multicolumn{2}{c}{$\bar{K}^*(1410)^0$} & \multicolumn{2}{c}{$K^*(1680)^-$} & \multicolumn{2}{c}{$K^*_2(1980)^-$} & \multicolumn{2}{c}{$(K\pi)_{\text{S-wave}}$} & \multicolumn{2}{c}{$(K\eta)_{\text{S-wave}}$} \\ \cline{3-16} 
                & & mag. & phase & mag. & phase & mag. & phase & mag. & phase & mag. & phase & mag. & phase & mag. & phase  \\ \hline
\multirow{2}{*}{$a_0(980)^+$} & mag. & $1$ & & & & & & & & & & & & &  \\
& phase & $-0.011$ & $1$ & & & & & & & & & & & &  \\
\multirow{2}{*}{$a_2(1320)^+$} & mag. & $-0.250$  & $0.188$  & $1$ & & & & & & & & & & &  \\
& phase & $-0.378$ & $-0.143$ & $-0.114$ & $1$ & & & & & & & & & &  \\
\multirow{2}{*}{$\bar{K}^*(1410)^0$} & mag. & $-0.699$ & $0.130$ & $-0.024$ & $0.511$ & $1$ & & & & & & & & & \\
& phase & $0.555$  & $-0.513$ & $-0.196$ & $-0.406$ &$-0.666$ & $1$  & & & & & & & & \\
\multirow{2}{*}{$K^*(1680)^-$} & mag. & $-0.516$ & $0.244$  & $0.231$  & $0.194$  &$0.596$  &$-0.240$ &$1$ & & & & & & &\\
& phase & $0.455$  & $-0.495$ & $-0.045$ & $-0.237$ &$-0.827$ &$0.707$ &$-0.572$ &$1$ & & & & & &  \\
\multirow{2}{*}{$K^*_2(1980)^-$} & mag. & $0.080$   & $-0.585$ & $0.092$  & $0.097$  &$-0.345$ &$0.469$ &$-0.302$ &$0.566$  &$1$ & & & & &\\
& phase & $0.208$  & $-0.606$ & $-0.514$ & $-0.164$ &$-0.232$ &$0.62$  &$-0.237$ &$0.441$  &$0.350$ &$1$ & & & & \\
\multirow{2}{*}{$(K\pi)_{\text{S-wave}}$} & mag. & $0.330$   & $-0.218$ & $0.095$  & $-0.483$ &$-0.582$ &$0.655$ &$-0.327$ &$0.502$  &$0.546$ &$0.246$ &$1$ & & &  \\
& phase & $-0.222$ & $0.649$  & $0.066$  & $0.143$  &$0.571$  &$-0.65$ &$0.521$  &$-0.812$ &$-0.652$ &$-0.461$ &$-0.609$ &$1$ & & \\
\multirow{2}{*}{$(K\eta)_{\text{S-wave}}$} & mag. & $0.493$  & $-0.503$ & $-0.189$ & $-0.326$ &$-0.687$ &$0.765$ &$-0.472$ &$0.804$  &$0.501$ &$0.528$ &$0.587$ &$-0.839$ &$1$ &  \\
& phase      & $0.375$  & $-0.451$ & $-0.194$ & $-0.089$ &$-0.615$ &$0.515$ &$-0.615$ &$0.790$ &$0.419$ &$0.376$ &$0.315$  &$-0.853$ &$0.768$ &$1$  \\ \hline
 & $B$        & $0.051$  & $-0.383$ & $-0.137$ & $0.170$  &$-0.157$ &$0.026$ &$-0.350$ &$0.371$ &$0.117$ &$0.345$ &$-0.375$ &$-0.199$ &$0.171$ &$0.488$   \\
LASS & $\phi_B$ & $0.210$  & $-0.538$ & $0.014$  & $-0.186$ &$-0.544$ &$0.607$ &$-0.438$ &$0.729$ &$0.654$ &$0.401$ &$0.611$  &$-0.858$ &$0.731$ &$0.687$  \\
of $K\pi$ & $\phi_R$ & $0.149$  & $-0.475$ & $-0.011$ & $-0.207$ &$-0.486$ &$0.561$ &$-0.373$ &$0.644$ &$0.550$  &$0.377$  &$0.689$  &$-0.842$ &$0.738$ &$0.715$\\
S-wave & $a$        & $0.302$  & $-0.653$ & $0.022$  & $-0.187$ &$-0.622$ &$0.677$ &$-0.534$ &$0.812$  &$0.702$  &$0.482$  &$0.637$  &$-0.934$ &$0.777$ &$0.739$ \\
& $r$        & $0.335$  & $-0.176$ & $0.159$  & $-0.258$ &$-0.425$ &$0.422$ &$-0.184$ &$0.301$  &$0.461$  &$0.086$  &$0.678$  &$-0.285$ &$0.255$ &$0.020$ \\ \hline
& $B$        & $-0.064$ & $-0.290$ & $-0.201$ & $0.268$  &$0.112$  &$0.065$ &$-0.026$ &$0.156$  &$0.152$  &$0.125$  &$-0.012$ &$-0.443$ &$0.440$ &$0.417$ \\
LASS & $\phi_B$ & $0.325$  & $-0.179$ & $0.163$  & $-0.599$ &$-0.566$ &$0.686$ &$-0.106$ &$0.405$  &$0.286$  &$0.257$  &$0.729$  &$-0.345$ &$0.369$ &$0.071$ \\
of $K\eta$ & $\phi_R$ & $-0.454$ & $0.441$  & $0.129$  & $0.329$  &$0.716$  &$-0.713$&$0.560$   &$-0.784$&$-0.451$ &$-0.414$&$-0.584$ &$0.872$  &$-0.786$&$-0.890$  \\
S-wave & $a$        & $-0.351$ & $0.528$  & $0.040$  & $0.258$  &$0.712$  &$-0.694$&$0.524$  &$-0.845$ &$-0.516$ &$-0.443$ &$-0.523$ &$0.830$  &$-0.736$ &$-0.779$  \\
& $r$        & $-0.329$ & $0.517$  & $0.024$  & $0.262$  &$0.702$  &$-0.681$&$0.519$  &$-0.830$ &$-0.512$ &$-0.435$ &$-0.520$ &$0.807$  &$-0.692$ &$-0.752$   \\
\hline \hline
\end{tabular}
  \end{lrbox}
  \scalebox{0.95}{\usebox{\tablebox}}
\end{centering}
\end{table}

To investigate the parameters of the Flatt\'e formulation of the $a_0(980)^{+}$ lineshape, the Dalitz fit based on the nominal model with free $g_{\pi^{\pm}\eta}$ is also performed and this yields $g_{\pi^{\pm}\eta}=0.596\pm0.008({\rm stat})$ GeV/$c^2$. This value is consistent with the measurement of BESIII, $0.607\pm0.011$ GeV/$c^2$~\cite{bes3_a0}. 
The significance of the $\pi\eta'$ contribution is tested and the results with floated $g_{\pi\eta'}$ and fixed $g_{\pi\eta'} = 0$ give $\Delta(-2\ln L)=102$ with $\Delta(\rm d.o.f)=1$, which indicates a $\pi\eta'$ contribution with 10.1$\sigma$ statistical significance. The fitted $g_{\pi\eta'}=0.408\pm0.018({\rm stat})$ GeV/$c^2$ is also consistent with the BESIII measurement of $g_{\pi\eta'}=0.424\pm0.050$ GeV/$c^2$~\cite{bes3_a0}.

\section{\boldmath Systematic uncertainties}{\label{systematic}}
The systematic uncertainties are divided into the uncertainties arising from the Dalitz model used in the fit and those from other sources. 
The model systematic uncertainties arise from the choice of individual components in the model, and from the parameterization of intermediate structures. 
The effective barrier radial parameter, $R$, is varied between 0 and 3.5 (GeV/$c$)$^{-1}$ for intermediate resonances, where the maximum value is chosen as the measured $R$ value for the narrowest resonance, the $K^{*}(892)$ ($R=3.0\pm0.5$ (GeV/$c$)$^{-1}$~\cite{PDG2018}), increased by its statistical error. 
Three coupling factors of the Flatt\'{e} function are varied within the quoted uncertainties, and the largest difference with respect to the nominal model is assigned as the systematic uncertainty due to this source. 
The masses and widths of intermediate resonances are varied within their uncertainties~\cite{PDG2018}. 
To account for the $K\pi$ and $K\eta$ S-wave components, the model used in the fit is modified by adding a wide resonance $\kappa$ described by a complex pole function~\cite{kappa2014} for a $K\pi$ S-wave, and $K^*_0(1950)^-$ described by RBW for a $K\eta$ S-wave. 
The nonsignificant resonance $a_0(1450)^+$ is added to evaluate the $\pi\eta$ S-wave component uncertainty. 
We also use a $\bar{K}^*(1680)^0$ resonance instead of a $\bar{K}^{*}(1410)^0$ contribution.

The systematic uncertainty due to the Dalitz distribution of combinatorial background is evaluated by (1) varying the $M$ sideband region within a shift of $\pm5$ MeV/$c^2$, and by (2) correcting the Dalitz distribution of experimental data in the $M$ sideband by the ratio of combinatorial background in the $M$ signal and sideband regions from generic MC. The larger difference is assigned as the systematic uncertainty due to the background distribution. 
The systematic uncertainty related to efficiency is estimated in two ways: (1) removing the correction for PID efficiency, and (2) shifting the $p^*(D^*)$ limit by $\pm0.05$ GeV/$c$ to consider possible discrepancy between MC and experimental data in $p^*(D^*)$ spectrum. These uncertainties are combined quadratically to give a systematic uncertainty due to efficiency. 
Comparing with the nominal fit model, the difference in the fit results when the signal fraction is varied by $\pm1\sigma$ (as determined from the $M$-$Q$ fit) is taken as the systematic uncertainty due to the uncertainty in the fraction of signal in the sample. 
A shift of the signal region by $\pm5$ MeV/$c^2$ in $M$ or $\pm0.1$ MeV/$c^2$ in $Q$ is applied to estimate the effect of the signal region selection. The larger difference in fit fraction is kept as the uncertainty due to this source. 
The uncertainty of multi-candidate selection is estimated by randomly selecting one of the multi-candidates as the best candidate instead of our nominal BCS method. 
The sources of systematic uncertainty considered are summarized in Tab.~\ref{tab:other_error}. Individual uncertainties are added in quadrature. 

\begin{table*}[!htbp]
\begin{center}
\caption{\label{tab:other_error} Sources of systematic uncertainties for each amplitude. For model systematic uncertainty: 1) Effective barrier radius $R$; 2) Flatt\'{e} coupling parameters $g_i$; 3) masses of resonances; 4) widths of resonances; 5) $K\pi$ S-wave uncertainty; 6) $K\eta$ S-wave uncertainty; 7) $\pi\eta$ S-wave uncertainty; 8) $\bar{K}^*(1680)^{0}$ instead of $\bar{K}^*(1410)^{0}$. For other sources: 9) signal fraction; 10) signal region; 11) background distribution; 12) efficiency variations; 13) best candidate selection for multi-candidates.}
\begin{tabular}{cc|ccccccccccc}
\hline \hline
\multicolumn{2}{c|}{Sources}  & ${\bar{K}^*(892)^0}$ & ${a_0(980)^+}$ & $K\pi$ S-wave & $K\eta$ S-wave & $a_2(1320)^+$ & $\bar{K}^*(1410)^{0}$ & $K^*(1680)^{-}$ & ${K_2^*(1980)^-}$ & $\sum FF$ \\ \hline
 & 1) & $^{+3.63}_{-2.64}$& $^{+3.26}_{-1.89}$& $^{+1.89}_{-1.34}$& $^{+0.28}_{-0.00}$& $^{+0.25}_{-0.10}$& $^{+1.37}_{-1.53}$& $^{+0.48}_{-0.01}$& $^{+0.85}_{-0.94}$& $^{+7.27}_{-3.47}$ \\
&2) & $^{+0.10}_{-0.09}$& $^{+2.11}_{-2.13}$& $^{+0.90}_{-1.01}$& $^{+0.22}_{-0.20}$& $^{+0.05}_{-0.05}$& $^{+0.83}_{-0.68}$& $^{+0.16}_{-0.13}$& $^{+0.14}_{-0.09}$& $^{+2.44}_{-2.40}$ \\ 
\multirow{2}{*}{model}&3) & $^{+0.17}_{-0.18}$& $^{+1.65}_{-1.64}$& $^{+0.80}_{-2.10}$& $^{+2.42}_{-1.01}$& $^{+0.02}_{-0.02}$& $^{+1.14}_{-1.81}$& $^{+0.16}_{-0.20}$& $^{+0.08}_{-0.11}$& $^{+1.50}_{-1.53}$ \\
\multirow{2}{*}{syst.} &4) & $^{+0.15}_{-0.39}$& $^{+0.69}_{-0.26}$& $^{+1.04}_{-0.96}$& $^{+1.43}_{-0.51}$& $^{+0.26}_{-0.12}$& $^{+1.30}_{-1.98}$& $^{+0.15}_{-0.08}$& $^{+0.13}_{-0.08}$& $^{+0.76}_{-0.48}$ \\
\multirow{2}{*}{(\%)}&5) & $+0.15$& $+0.62$& $+1.01$& $+1.47$& $+0.03$& $-0.57$& $-0.23$& $-0.07$& $+2.41$ \\
&6) & $+0.06$& $-0.06$& $+0.65$& $-0.29$& $+0.00$& $-0.19$& $-0.13$& $+0.02$& $+1.07$ \\
 &7) & $-0.09$& $-0.20$& $-0.33$& $+0.15$& $+0.00$& $+0.23$& $+0.00$& $+0.07$& $-0.16$ \\ 
&8) & $-0.43$& $+0.74$& $-0.05$& $-0.22$& $+0.03$& $+0.16$& $+0.17$& $-0.23$& $+0.17$ \\ 
&Total& $^{+3.64}_{-2.71}$& $^{+4.38}_{-3.30}$& $^{+2.75}_{-2.87}$& $^{+3.20}_{-1.21}$& $^{+0.37}_{-0.17}$& $^{+2.37}_{-3.22}$& $^{+0.58}_{-0.36}$& $^{+0.88}_{-0.98}$& $^{+8.28}_{-4.52}$ \\ \hline
&9) & $^{+0.03}_{-0.06}$& $^{+0.05}_{-0.06}$& $^{+0.15}_{-0.00}$& $^{+0.06}_{-0.00}$& $^{+0.01}_{-0.00}$& $^{+0.05}_{-0.06}$& $^{+0.01}_{-0.00}$& $^{+0.01}_{-0.00}$& $^{+0.14}_{-0.01}$ \\
\multirow{2}{*}{other}&10)& $^{+0.10}_{-0.19}$& $^{+0.00}_{-0.46}$& $^{+0.80}_{-0.50}$& $^{+0.21}_{-0.22}$& $^{+0.05}_{-0.03}$& $^{+0.33}_{-0.15}$& $^{+0.06}_{-0.10}$& $^{+0.04}_{-0.02}$& $^{+0.54}_{-0.65}$ \\
\multirow{2}{*}{syst.}&11)& $^{+0.06}_{-0.22}$& $^{+0.54}_{-0.22}$& $^{+0.35}_{-0.16}$& $^{+0.14}_{-0.16}$& $^{+0.03}_{-0.01}$& $^{+0.44}_{-0.22}$& $^{+0.07}_{-0.02}$& $^{+0.03}_{-0.05}$& $^{+0.62}_{-0.20}$ \\
\multirow{2}{*}{(\%)}&12)& $^{+0.21}_{-0.00}$& $^{+0.04}_{-0.04}$& $^{+0.16}_{-0.04}$& $^{+0.14}_{-0.00}$& $^{+0.01}_{-0.00}$& $^{+0.00}_{-0.21}$& $^{+0.05}_{-0.00}$& $^{+0.01}_{-0.00}$& $^{+0.22}_{-0.05}$ \\
&13)& $-0.39$& $+1.48$& $+1.16$& $+0.71$& $-0.02$& $-1.57$& $+0.01$& $+0.00$& $+1.29$ \\ 
&Total& $^{+0.24}_{-0.49}$& $^{+1.58}_{-0.51}$& $^{+1.47}_{-0.53}$& $^{+0.77}_{-0.27}$& $^{+0.06}_{-0.04}$& $^{+0.55}_{-1.61}$& $^{+0.11}_{-0.10}$& $^{+0.05}_{-0.05}$& $^{+1.55}_{-0.68}$ \\
\hline \hline
\end{tabular}
\end{center}
\end{table*}

\section{Further study and discussion}{\label{sec:discussion}}
In this section, we present further discussion of the Dalitz fit results shown in Tab.~\ref{tab:DPfitexp} and of our measured branching fraction $\mathcal{B}(D^0\to\Km\pip\eta)=(1.973\pm0.009{\rm(stat)}\pm0.079{\rm(syst)}\pm0.018{\rm (\mathcal{B}_{PDG})})\%$. 
\begin{itemize}
\item $D^0\to \bar{K}^{*}(892)^0\eta$ decay: 
we determine $\mathcal{B}(D^0\to[\bar{K}^{*}(892)^0\to\Km\pip]\eta)=(0.94\pm0.03{\rm(stat)}^{+0.08}_{-0.07}{\rm(syst)}\pm0.01{\rm(\mathcal{B})})\%$. 
Using $\mathcal{B}(\bar{K}^{*}(892)^0\to\Km\pip)=(66.503\pm0.014)\%$~\cite{PDG2018}, we find $\mathcal{B}(D^0\to\bar{K}^{*}(892)\eta)=(1.41\pm0.04{\rm(stat)}^{+0.12}_{-0.11}{\rm(syst)}\pm0.01{\rm(\mathcal{B}_{\text{PDG}})})\%=(1.41^{+0.13}_{-0.12})\%$, 
which is consistent with, and more precise than, the current world average $(1.02\pm0.30)\%$~\cite{PDG2018}. It deviates from the theoretical predictions of (0.51-0.92)\%~\cite{Cheng:2010ry,Li:2012cfa,Li:2013xsa} with a significance of more than $3\sigma$. 

\item $K^{*}(1680)\to K\eta$ decay: 
we determine $\mathcal{B}(D^0\to[K^{*}(1680)^-\to\Km\eta]\pip)=(2.11\pm0.32{\rm(stat)}^{+1.16}_{-0.72}{\rm(syst)}\pm0.02{\rm(\mathcal{B})})\times10^{-4}$.
Using $\mathcal{B}(D^0\to [K^{*}(1680)^-\to\Km\piz]\pip)=(0.19\pm0.07)\%$~\cite{PDG2018} and $\mathcal{B}(K^{*}(1680)^-\to\Km\piz)=(12.90\pm0.83)\%$~\cite{PDG2018}, the branching fraction of $D^0\to K^{*}(1680)^-\pip$ is $(1.47\pm0.55)\%$.
Thus, one obtains $\mathcal{B}(K^{*}(1680)^-\to\Km\eta)=(1.44\pm0.21{\rm(stat)}^{+0.79}_{-0.49}{\rm(syst)}\pm0.54{\rm(\mathcal{B}_{\text{PDG}})})\%$, where the uncertainties are respectively statistical, systematic, and due to the branching fraction uncertainties in Ref.~\cite{PDG2018}. 
Finally, the relative branching ratio of $K^{*}(1680)^-$ to $\Km\eta$ and $\Km\piz$~\cite{PDG2018} is measured to be $0.11\pm0.02{\rm(stat)}^{+0.06}_{-0.04}{\rm(syst)}\pm0.04{\rm(\mathcal{B}_{\text{PDG}})}$, which is not consistent with theoretical predictions ($\approx1.0$) under the assumption that $K^{*}(1680)$ is a pure $1^{3}D_1$ state~\cite{Barnes:2002mu,Pang:2017dlw}. 
This ratio is comparable to $\mathcal{B}(K_0^{*}(1430)\to K\eta)/\mathcal{B}(K_0^{*}(1430)\to K\pi)=0.09^{+0.03}_{-0.04}$~\cite{PDG2018}, which is consistent with the theoretical prediction assuming that $K^{*}(1430)$ is a $1^{3}P_0$ state~\cite{Barnes:2002mu,Pang:2017dlw}.

\item $K_2^{*}(1980)\to K\eta$ decay: 
we measure for the first time $\mathcal{B}(D^0\to[K_2^{*}(1980)^-\to\Km\eta]\pip)=(2.2\pm0.2{\rm(stat)}^{+1.7}_{-1.9}{\rm(syst)}\pm0.0{\rm(\mathcal{B}_{\text{PDG}})})\times10^{-4}=(2.2^{+1.7}_{-1.9})\times10^{-4}$, which is strongly suppressed due to a limit of the phase-space region and yet allowed due to a large width of $K^{*}_2(1980)$.

\end{itemize}

\section{\boldmath conclusion}{\label{conclusions}}
In summary, using 953 ${\rm fb^{-1}}$ of data collected by the Belle detector, a Dalitz plot analysis of $\Dz\to\Km\pip\eta$ is performed. The DP is well represented by a combination of significant quasi-two-body decay channels with six intermediate resonances:  $\bar{K}^{*}(892)^0$, $a_0(980)^+$, $a_2(1320)^+$, $\bar{K}^{*}(1410)^0$, $K^{*}(1680)^-$, $K_2^{*}(1980)^-$, and two S-wave components of $K\pi$ and $K\eta$. The fit fraction of each component is given in Tab.~\ref{tab:DPfitexp}. 
The dominant contributions to the decay amplitude arise from $\bar{K}^{*}(892)^{0}$, $a_0(980)^{+}$ and the $K\pi$ S-wave component. 
The $K\eta$ S-wave component, including $K_0^{*}(1430)^{-}$, is observed with a statistical significance of more than $30\sigma$, and the decays $K^{*}(1680)^{-}\to\Km\eta$ and $K^{*}_2(1980)^{-}\to\Km\eta$ are observed for the first time and have statistical significances of $16\sigma$ and $17\sigma$, respectively. 

We measure the ratio of the branching fractions, $\frac{\mathcal{B}(D^0\to\Km\pip\eta)}{\mathcal{B}(D^0\to\Km\pip)}=0.500\pm0.002{\rm(stat)}\pm0.020{\rm(syst)}\pm0.003{\rm (\mathcal{B}_{PDG})}$ for the first time. 
The relative branching ratio $\frac{\mathcal{B}(K^*(1680)^-\to\Km\eta)}{\mathcal{B}(K^*(1680)^-\to\Km\piz)}$ is determined to be $0.11\pm0.02{\rm(stat)}^{+0.06}_{-0.04}{\rm(syst)}\pm0.04{\rm(\mathcal{B}_{\text{PDG}})}$. This is not consistent with the theoretical prediction under an assumption of a pure $1^{3}D_1$ state~\cite{Barnes:2002mu,Pang:2017dlw}. 
We also determine the product of branching fraction $\mathcal{B}(D^0\to[K_2^{*}(1980)^-\to\Km\eta]\pip)=(2.2^{+1.7}_{-1.9})\times10^{-4}$.
For $a_0(980)^+$, we confirm the $\pi\eta^{\prime}$ contribution in the three-channel Flatt\'{e} model with a statistical significance of $10.1\sigma$. 
We have also determined the branching fraction $\mathcal{B}(D^0\to\bar{K}^{*}(892)^0\eta)=(1.41^{+0.13}_{-0.12})\%$, which is consistent with, and more precise than, the current world average of $(1.02\pm0.30)\%$~\cite{PDG2018}. It deviates from the various theoretical predictions of (0.51-0.92)\%~\cite{Cheng:2010ry,Li:2012cfa,Li:2013xsa} with a significance of more than $3\sigma$.

\section{\boldmath ACKNOWLEDGMENTS}
We warmly thank Prof. Cheng-Qun Pang, Prof. Fu-Sheng Yu and Dr. Zhen-Tian Sun for interesting discussions.
We thank the KEKB group for the excellent operation of the
accelerator; the KEK cryogenics group for the efficient
operation of the solenoid; and the KEK computer group, and the Pacific Northwest National
Laboratory (PNNL) Environmental Molecular Sciences Laboratory (EMSL)
computing group for strong computing support; and the National
Institute of Informatics, and Science Information NETwork 5 (SINET5) for
valuable network support.  We acknowledge support from
the Ministry of Education, Culture, Sports, Science, and
Technology (MEXT) of Japan, the Japan Society for the 
Promotion of Science (JSPS), and the Tau-Lepton Physics 
Research Center of Nagoya University; 
the Australian Research Council including grants
DP180102629, 
DP170102389, 
DP170102204, 
DP150103061, 
FT130100303; 
Austrian Science Fund (FWF);
the National Natural Science Foundation of China under Contracts
No.~11435013,  
No.~11475187,  
No.~11521505,  
No.~11575017,  
No.~11675166,  
No.~11705209;  
Key Research Program of Frontier Sciences, Chinese Academy of Sciences (CAS), Grant No.~QYZDJ-SSW-SLH011; 
the  CAS Center for Excellence in Particle Physics (CCEPP); 
the Shanghai Pujiang Program under Grant No.~18PJ1401000;  
the Ministry of Education, Youth and Sports of the Czech
Republic under Contract No.~LTT17020;
the Carl Zeiss Foundation, the Deutsche Forschungsgemeinschaft, the
Excellence Cluster Universe, and the VolkswagenStiftung;
the Department of Science and Technology of India; 
the Istituto Nazionale di Fisica Nucleare of Italy; 
National Research Foundation (NRF) of Korea Grant
Nos.~2016R1\-D1A1B\-01010135, 2016R1\-D1A1B\-02012900, 2018R1\-A2B\-3003643,
2018R1\-A6A1A\-06024970, 2018R1\-D1A1B\-07047294, 2019K1\-A3A7A\-09033840,
2019R1\-I1A3A\-01058933;
Radiation Science Research Institute, Foreign Large-size Research Facility Application Supporting project, the Global Science Experimental Data Hub Center of the Korea Institute of Science and Technology Information and KREONET/GLORIAD;
the Polish Ministry of Science and Higher Education and 
the National Science Center;
the Ministry of Science and Higher Education of the Russian Federation, Agreement 14.W03.31.0026; 
University of Tabuk research grants
S-1440-0321, S-0256-1438, and S-0280-1439 (Saudi Arabia);
the Slovenian Research Agency;
Ikerbasque, Basque Foundation for Science, Spain;
the Swiss National Science Foundation; 
the Ministry of Education and the Ministry of Science and Technology of Taiwan;
and the United States Department of Energy and the National Science Foundation.

\end{document}

%% file: author.tex
\noaffiliation
\affiliation{University of the Basque Country UPV/EHU, 48080 Bilbao}
\affiliation{Beihang University, Beijing 100191}
\affiliation{University of Bonn, 53115 Bonn}
\affiliation{Brookhaven National Laboratory, Upton, New York 11973}
\affiliation{Budker Institute of Nuclear Physics SB RAS, Novosibirsk 630090}
\affiliation{Faculty of Mathematics and Physics, Charles University, 121 16 Prague}
\affiliation{Chonnam National University, Gwangju 61186}
\affiliation{University of Cincinnati, Cincinnati, Ohio 45221}
\affiliation{Deutsches Elektronen--Synchrotron, 22607 Hamburg}
\affiliation{Department of Physics, Fu Jen Catholic University, Taipei 24205}
\affiliation{Key Laboratory of Nuclear Physics and Ion-beam Application (MOE) and Institute of Modern Physics, Fudan University, Shanghai 200443}
\affiliation{Justus-Liebig-Universit\"at Gie\ss{}en, 35392 Gie\ss{}en}
\affiliation{Gifu University, Gifu 501-1193}
\affiliation{SOKENDAI (The Graduate University for Advanced Studies), Hayama 240-0193}
\affiliation{Gyeongsang National University, Jinju 52828}
\affiliation{Department of Physics and Institute of Natural Sciences, Hanyang University, Seoul 04763}
\affiliation{University of Hawaii, Honolulu, Hawaii 96822}
\affiliation{High Energy Accelerator Research Organization (KEK), Tsukuba 305-0801}
\affiliation{J-PARC Branch, KEK Theory Center, High Energy Accelerator Research Organization (KEK), Tsukuba 305-0801}
\affiliation{Forschungszentrum J\"{u}lich, 52425 J\"{u}lich}
\affiliation{IKERBASQUE, Basque Foundation for Science, 48013 Bilbao}
\affiliation{Indian Institute of Science Education and Research Mohali, SAS Nagar, 140306}
\affiliation{Indian Institute of Technology Bhubaneswar, Satya Nagar 751007}
\affiliation{Indian Institute of Technology Hyderabad, Telangana 502285}
\affiliation{Indian Institute of Technology Madras, Chennai 600036}
\affiliation{Indiana University, Bloomington, Indiana 47408}
\affiliation{Institute of High Energy Physics, Chinese Academy of Sciences, Beijing 100049}
\affiliation{Institute of High Energy Physics, Vienna 1050}
\affiliation{Institute for High Energy Physics, Protvino 142281}
\affiliation{INFN - Sezione di Napoli, 80126 Napoli}
\affiliation{INFN - Sezione di Torino, 10125 Torino}
\affiliation{Advanced Science Research Center, Japan Atomic Energy Agency, Naka 319-1195}
\affiliation{J. Stefan Institute, 1000 Ljubljana}
\affiliation{Institut f\"ur Experimentelle Teilchenphysik, Karlsruher Institut f\"ur Technologie, 76131 Karlsruhe}
\affiliation{Kennesaw State University, Kennesaw, Georgia 30144}
\affiliation{King Abdulaziz City for Science and Technology, Riyadh 11442}
\affiliation{Department of Physics, Faculty of Science, King Abdulaziz University, Jeddah 21589}
\affiliation{Kitasato University, Sagamihara 252-0373}
\affiliation{Korea Institute of Science and Technology Information, Daejeon 34141}
\affiliation{Korea University, Seoul 02841}
\affiliation{Kyungpook National University, Daegu 41566}
\affiliation{LAL, Univ. Paris-Sud, CNRS/IN2P3, Universit\'{e} Paris-Saclay, Orsay 91898}
\affiliation{\'Ecole Polytechnique F\'ed\'erale de Lausanne (EPFL), Lausanne 1015}
\affiliation{P.N. Lebedev Physical Institute of the Russian Academy of Sciences, Moscow 119991}
\affiliation{Faculty of Mathematics and Physics, University of Ljubljana, 1000 Ljubljana}
\affiliation{Ludwig Maximilians University, 80539 Munich}
\affiliation{Luther College, Decorah, Iowa 52101}
\affiliation{Malaviya National Institute of Technology Jaipur, Jaipur 302017}
\affiliation{University of Maribor, 2000 Maribor}
\affiliation{Max-Planck-Institut f\"ur Physik, 80805 M\"unchen}
\affiliation{School of Physics, University of Melbourne, Victoria 3010}
\affiliation{University of Mississippi, University, Mississippi 38677}
\affiliation{University of Miyazaki, Miyazaki 889-2192}
\affiliation{Moscow Physical Engineering Institute, Moscow 115409}
\affiliation{Moscow Institute of Physics and Technology, Moscow Region 141700}
\affiliation{Graduate School of Science, Nagoya University, Nagoya 464-8602}
\affiliation{Universit\`{a} di Napoli Federico II, 80055 Napoli}
\affiliation{Nara Women's University, Nara 630-8506}
\affiliation{National Central University, Chung-li 32054}
\affiliation{National United University, Miao Li 36003}
\affiliation{Department of Physics, National Taiwan University, Taipei 10617}
\affiliation{H. Niewodniczanski Institute of Nuclear Physics, Krakow 31-342}
\affiliation{Nippon Dental University, Niigata 951-8580}
\affiliation{Niigata University, Niigata 950-2181}
\affiliation{Novosibirsk State University, Novosibirsk 630090}
\affiliation{Osaka City University, Osaka 558-8585}
\affiliation{Pacific Northwest National Laboratory, Richland, Washington 99352}
\affiliation{Panjab University, Chandigarh 160014}
\affiliation{Peking University, Beijing 100871}
\affiliation{University of Pittsburgh, Pittsburgh, Pennsylvania 15260}
\affiliation{Punjab Agricultural University, Ludhiana 141004}
\affiliation{Theoretical Research Division, Nishina Center, RIKEN, Saitama 351-0198}
\affiliation{Department of Modern Physics and State Key Laboratory of Particle Detection and Electronics, University of Science and Technology of China, Hefei 230026} 
\affiliation{Showa Pharmaceutical University, Tokyo 194-8543}
\affiliation{Soochow University, Suzhou 215006}
\affiliation{Soongsil University, Seoul 06978}
\affiliation{University of South Carolina, Columbia, South Carolina 29208}
\affiliation{Sungkyunkwan University, Suwon 16419}
\affiliation{School of Physics, University of Sydney, New South Wales 2006}
\affiliation{Department of Physics, Faculty of Science, University of Tabuk, Tabuk 71451}
\affiliation{Tata Institute of Fundamental Research, Mumbai 400005}
\affiliation{Department of Physics, Technische Universit\"at M\"unchen, 85748 Garching}
\affiliation{Toho University, Funabashi 274-8510}
\affiliation{Department of Physics, Tohoku University, Sendai 980-8578}
\affiliation{Earthquake Research Institute, University of Tokyo, Tokyo 113-0032}
\affiliation{Department of Physics, University of Tokyo, Tokyo 113-0033}
\affiliation{Tokyo Institute of Technology, Tokyo 152-8550}
\affiliation{Tokyo Metropolitan University, Tokyo 192-0397}
\affiliation{Utkal University, Bhubaneswar 751004}
\affiliation{Virginia Polytechnic Institute and State University, Blacksburg, Virginia 24061}
\affiliation{Wayne State University, Detroit, Michigan 48202}
\affiliation{Yamagata University, Yamagata 990-8560}
\affiliation{Yonsei University, Seoul 03722}

  \author{Y.~Q.~Chen}\affiliation{Department of Modern Physics and State Key Laboratory of Particle Detection and Electronics, University of Science
and Technology of China, Hefei 230026} 
  \author{L.~K.~Li$^{*}$}\affiliation{Institute of High Energy Physics, Chinese Academy of Sciences, Beijing 100049}\affiliation{University of Cincinnati, Cincinnati, Ohio 45221} 
  \author{W.~B.~Yan}\affiliation{Department of Modern Physics and State Key Laboratory of Particle Detection and Electronics, University of Science
and Technology of China, Hefei 230026} 

  \author{I.~Adachi}\affiliation{High Energy Accelerator Research Organization (KEK), Tsukuba 305-0801}\affiliation{SOKENDAI (The Graduate University for Advanced Studies), Hayama 240-0193} 
  \author{H.~Aihara}\affiliation{Department of Physics, University of Tokyo, Tokyo 113-0033} 
  \author{S.~Al~Said}\affiliation{Department of Physics, Faculty of Science, University of Tabuk, Tabuk 71451}\affiliation{Department of Physics, Faculty of Science, King Abdulaziz University, Jeddah 21589} 
  \author{D.~M.~Asner}\affiliation{Brookhaven National Laboratory, Upton, New York 11973} 
  \author{H.~Atmacan}\affiliation{University of Cincinnati, Cincinnati, Ohio 45221} 
  \author{V.~Aulchenko}\affiliation{Budker Institute of Nuclear Physics SB RAS, Novosibirsk 630090}\affiliation{Novosibirsk State University, Novosibirsk 630090} 
  \author{T.~Aushev}\affiliation{Moscow Institute of Physics and Technology, Moscow Region 141700} 
% ...
\collaboration{The Belle Collaboration}

%% file: main.bbl
\begin{thebibliography}{99}
\bibitem{Cheng:2010ry}
H.~Y. Cheng and C.~W. Chiang, \href{https://journals.aps.org/prd/abstract/10.1103/PhysRevD.81.074021}{Phys. Rev. D {\bf 81}, 074021 (2010)}. 

\bibitem{Li:2012cfa}
H.~N. Li, C.~D. L\"u, and F.~S. Yu, \href{https://journals.aps.org/prd/abstract/10.1103/PhysRevD.86.036012}{Phys. Rev. D {\bf 86}, 036012 (2012)}. 

\bibitem{Li:2013xsa}
Q. Qin, H.~N. Li, C.~D. L\"u, and F.~S. Yu, \href{https://link.aps.org/doi/10.1103/PhysRevD.89.054006}{Phys. Rev. D {\bf 89}, 054006 (2014)}.

\bibitem{Dalitz} 
R.~H. Dalitz, \href{https://www.tandfonline.com/doi/abs/10.1080/14786441008520365}{Philos. Mag. {\bf 44}, 1068 (1953)}.

\bibitem{Rubin:2004cq} 
P.~Rubin {\it et al.} (CLEO Collaboration), \href{http://journals.aps.org/prl/abstract/10.1103/PhysRevLett.93.111801}{Phys. Rev. Lett. {\bf 93}, 111801 (2004)}.

\bibitem{PDG2018} 
M. Tanabashi {\it et al.} (Particle Data Group), \href{https://journals.aps.org/prd/abstract/10.1103/PhysRevD.98.030001}{Phys. Rev. D {\bf 98}, 030001 (2018)}.

\bibitem{BABAR_Lees2014} 
J.~P.~Lees {\it et al.} (BABAR Collaboration), \href{http://journals.aps.org/prd/abstract/10.1103/PhysRevD.89.112004}{Phys. Rev. D {\bf 89}, 112004 (2014)}.

\bibitem{Barnes:2002mu} 
T. Barnes, N. Black, and P.~R.~Page, \href{https://journals.aps.org/prd/abstract/10.1103/PhysRevD.68.054014}{Phys. Rev. D {\bf 68}, 054014 (2003)}.

\bibitem{Pang:2017dlw}
C.~Q.~Pang, J.~Z.~Wang, X. Liu, {\it et al.} \href{https://doi.org/10.1140/epjc/s10052-017-5434-0}{Eur. Phys. J. C (2017) 77: 861}. 

\bibitem{bes3_a0} 
M.~Ablikim {\it et al.} (BESIII Collaboration), \href{https://journals.aps.org/prd/abstract/10.1103/PhysRevD.95.032002}{Phys. Rev. D {\bf 95}, 032002 (2017)}.

\bibitem{ddbarmixngLHCb}
R. Aaij {\it et al.} (LHCb Collaboration), \href{https://journals.aps.org/prl/abstract/10.1103/PhysRevLett.110.101802}{Phys. Rev. Lett. {\bf 110}, 101802 (2013)}.

\bibitem{bib:BelleII}
E. Kou {\it et al.}, \href{https://academic.oup.com/ptep/article/2019/12/123C01/5685006}{Prog. Theor. Exp. Phys. {\bf 2019} (2019) 123C01}.

\bibitem{bib:conjugated}
The inclusion of charge-conjugate reactions is implied here and throughout this paper.

\bibitem{BelleDetector} 
A.~Abashian {\it et al.} (Belle Collaboration), \href{http://www.sciencedirect.com/science/article/pii/S0168900201020137}{Nucl. Instrum. Methods 
 Phys. Res. Sect. A {\bf 479}, 117 (2002)}; also see detector section in
 J.~Brodzicka {\it et al.}, \href{http://ptep.oxfordjournals.org/content/2012/1/04D001.abstract}{Prog. Theor. Exp. Phys. {\bf 2012}, 04D001 (2012)}.

\bibitem{KEKB} 
S.~Kurokawa and E.~Kikutani, \href{https://www.sciencedirect.com/science/article/pii/S0168900202017709}{Nucl. Instrum. Methods Phys. Res. Sect. A {\bf 499}, 1 (2003)}, and other papers included in this Volume; T.Abe {\it et al.}, \href{https://academic.oup.com/ptep/article/2013/3/03A001/1556871}{Prog. Theor. Exp. Phys. {\bf 2013}, 03A001 (2013)} and references therein.

\bibitem{whyTag}
$D^0$ meson flavor tagging is used to suppress the combinatorial background and doubly-Cabibbo-suppressed decays.

\bibitem{bib:PID}
E.~Nakano, \href{https://www.sciencedirect.com/science/article/pii/S0168900202015103}{Nucl. Instrum. Methods Phys. Res. Sect. A {\bf 494}, 402 (2002)}.

\bibitem{evtgen} 
D.~J.~Lange, The {\scshape{EVTGEN}} particle decay simulation package, \href{http://www.sciencedirect.com/science/article/pii/S0168900201000894}{{Nucl. Instrum. Methods Phys. Res. Sect.} {\bf A462}, 152 (2001)}.

\bibitem{jetset} 
T.~Sj\"ostrand, High-energy physics event generation with {\scshape{Pythia}} 5.7 and JetSet 7.4, \href{https://doi.org/10.1016/0010-4655(94)90132-5}{Comput. Phys. Commun. {\bf 82}, 74 (1994)}.

\bibitem{geant3} 
R.~Brun {\it et al.}, GEANT 3.21, CERN Report No. DD/EE/84-1, 1984.

\bibitem{photons} 
E.~Barberio and Z.~W\k{a}s, \href{https://doi.org/10.1016/0010-4655(94)90074-4}{Comput. Phys. Commun. {\bf 79}, 291 (1994)}.

\bibitem{Cruijff}
The bifurcated Cruijff function is a centered Gaussian with asymmetry resolution and non-Gaussian tails: $f(x) = \exp((x-\mu)^2/(2\sigma_{L,R}^2 + \alpha_{L,R}(x-\mu)^2))$.

\bibitem{bilinear} 
W.~H. Press, S.~A. Teukolsky, W.~T. Vetterling and B.P. Flannery, \href{https://archive.org/details/numericalrecipes0865unse/page/n5}{{\it Numerical recipes in C: the art of scientific computing} (2nd ed.)}. New York, NY, USA: Cambridge University Press. pp. 123-128. 

\bibitem{isobar model} R.~M.~Sternheimer and S. J. Lindenbaum, \href{https://journals.aps.org/pr/pdf/10.1103/PhysRev.123.333}{Phys. Rev. {\bf 123}, 333-376 (1961)}.

\bibitem{Zemach} Ch.~Zemach,  \href{http://inspirehep.net/record/4753?ln=en}{Phys. Rev. {\bf 133}, B1201 (1964)}; Ch.~Zemach,  \href{https://journals.aps.org/pr/pdf/10.1103/PhysRev.140.B97}{Phys. Rev. {\bf 140}, B109 (1965)}.

\bibitem{Kpipi0_CLEO} 
S.~Kopp {\it et al.} (CLEO Collaboration), \href{https://journals.aps.org/prd/abstract/10.1103/PhysRevD.63.092001}{Phys. Rev. D {\bf 63}, 092001 (2001)}

\bibitem{bib:BWfactor}
J. Blatt and V. Weisskopf, {\it Theoretical Nuclear Physics}, (\href{https://www.springer.com/gp/book/9781461299615}{Springer-Verlag, 1979}).

\bibitem{bib:BWfactor2}
F. von Hippel and C. Quigg, \href{https://journals.aps.org/prd/abstract/10.1103/PhysRevD.5.624}{Phys. Rev. D {\bf 5}, 624 (1972)}.

\bibitem{PBF} 
Ed.~A.~J.~Bevan, B.~Golob, Th.~Mannel, S.~Prell, and B.~D.~Yabsley, \href{http://arxiv.org/abs/1406.6311}{Eur. Phys. J. C {\bf 74}, (2014) 3026}.

\bibitem{LASS2} 
R.~Aaij {\it et al.} (LHCb Collaboration), \href{https://journals.aps.org/prd/abstract/10.1103/PhysRevD.93.052018}{Phys. Rev. D {\bf 93}, 052018 (2016)}.

\bibitem{bib:resonances}
The intermediate resonances from \cite{PDG2018} are checked in this Dalitz analysis: $\kappa$, $K^{*}(892)^0$, $K^{*}(1410)^{0}$, $K_0^{*}(1430)^{0}$, and $K_2^{*}(1430)^0$ for $K^-\pi^+$ mass spectrum; $K^{*}(1410)^{-}$, $K_0^{*}(1430)^{-}$, $K_2^{*}(1430)^-$, $K^{*}(1680)^{-}$, $K_3^{*}(1780)^{-}$, $K^{*}(1950)^-$, and $K_2^{*}(1980)^-$ for $K^-\eta$ mass spectrum; $a_0(980)^+$, $a_2(1320)^+$, and $a_0(1450)^+$ for $\pi^+\eta$ mass spectrum.

\bibitem{kappa2014} 
E.~M.~Aitala {\it et al.} (Fermilab E791 Collaboration), \href{https://journals.aps.org/prl/abstract/10.1103/PhysRevLett.89.121801}{Phys. Rev. Lett. {\bf 89}, 121801 (2002)}; 
M.~Ablikim {\it et al.} (BESIII Collaboration),  \href{https://journals.aps.org/prd/pdf/10.1103/PhysRevD.89.052001}{Phys. Rev. D {\bf 89}, 052001 (2014)}.

\end{thebibliography}
